\documentclass{anstrans}
\usepackage{microtype}
\usepackage[utf8]{inputenc}
\usepackage{amsmath}
\usepackage{amsfonts}
\usepackage{amssymb}
\usepackage{graphicx} 
\usepackage{floatrow}
\usepackage[dvipsnames]{xcolor}
\usepackage{makecell}
\usepackage[per-mode=symbol,round-precision=3,scientific-notation=false,range-phrase = \text{--}]{siunitx}
\usepackage{booktabs} 
\usepackage[hyperindex,breaklinks]{hyperref}

\usepackage{floatrow}
\usepackage{xspace}
\newcommand*{\eg}{e.g.,\@\xspace}
\newcommand*{\ie}{i.e.,\@\xspace}

\newcommand*{\vs}{vs.\@\xspace}

\makeatletter
\newcommand*{\etc}{%
    \@ifnextchar{.}%
        {etc}%
        {etc.\@\xspace}%
}
\makeatother

\usepackage[nameinlink,capitalize]{cleveref}
\DeclareUnicodeCharacter{025B}{\ensuremath{\varepsilon}}

\graphicspath{{figures/}}

\usepackage{listings}

\definecolor{bkgd}{RGB}{240,242,246}
\definecolor{ceruleanblue}{rgb}{0.16, 0.32, 0.75}
\definecolor{orange-red}{rgb}{1.0, 0.27, 0.0}
\definecolor{anotherblue}{RGB}{37,92,243}
\definecolor{blackblue}{RGB}{46,60,85}
\definecolor{goldyellow}{RGB}{199,146,12}
\lstdefinestyle{altstyle2}{
    backgroundcolor=\color{bkgd},
    basicstyle=\ttfamily\footnotesize\color{blackblue},
    breakatwhitespace=false,
    breaklines=true,
    captionpos=b,
    commentstyle=\color{goldyellow},
    keepspaces=true,
    keywordstyle=\color{orange-red},
    language=Python,
    numbersep=5pt,
    numberstyle=\tiny\color{ceruleanblue},
    showspaces=false,
    showstringspaces=false,
    showtabs=false,
    stringstyle=\color{anotherblue},
    tabsize=2
}
\newcommand{\pkg}[1]{\texttt{#1}}

\usepackage[scaled=0.86]{helvet}
\newcommand*{\helvet}{\fontfamily{phv}\selectfont}
\newcommand{\htxt}[1]{{\helvet{#1}}}

\usepackage[abs]{overpic}
\usepackage{multirow}

\definecolor{hypred}{RGB}{236,34,39}

\title{Deep Surrogate Models for Multi-dimensional Regression of Reactor Power}
\author{Akshay J.\@\xspace Dave$^{*}$, Jarod Wilson$^{*}$, Kaichao Sun$^{*}$}

\institute{
$^{*}$Nuclear Reactor Laboratory, Massachusetts Institute of Technology, Cambridge, MA, akshayjd@mit.edu
}

\begin{document}
\section{Introduction}

There is renewed interest in developing small modular reactors and micro-reactors.
Innovation is necessary in both construction and operation methods of these reactors to be financially attractive \cite{petti2018future}.
For construction, methods such as additive manufacturing \cite{ornl3dprint} is under active development.
For operation, an area of interest is the development of fully autonomous reactor control \cite{Wood2017}.
Significant efforts are necessary to demonstrate an autonomous control framework for a nuclear system, while adhering to established safety criteria.
For \textit{critical} reactors of interest, the latter precludes implementing such a framework.
Therefore, our group has proposed and received support for demonstration of an autonomous framework on a subcritical system: the MIT Graphite Exponential Pile (MGEP)\footnote{Details on the MGEP available at \cite{Gale2018}.}.

\begin{figure}
\includegraphics[width=\linewidth]{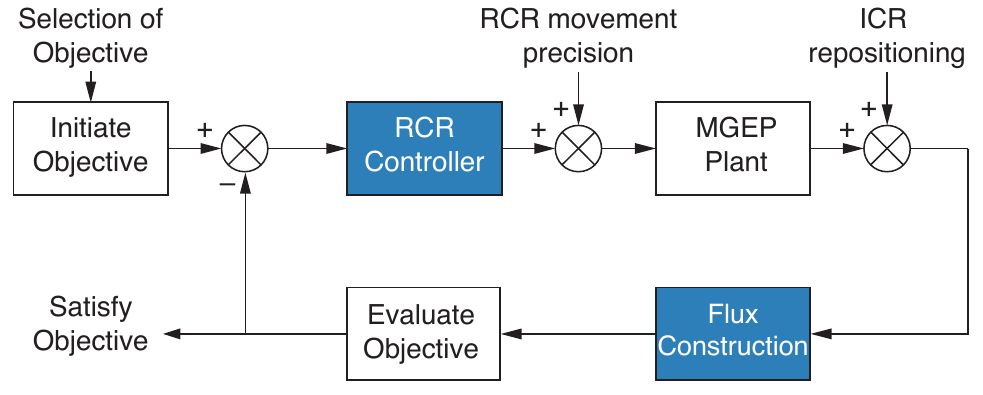}
\caption{Control system layout. 
	Blue boxes indicate processes where a neural network surrogate can be deployed. 
	RCR: Reacting control rod (controlled); ICR: Initiating control rod (unknown pertubation).}
\label{fig:system}
\end{figure}

The autonomous system under development aims to incorporate a surrogate model.
Why is a surrogate model necessary?
The preliminary system layout is presented in \cref{fig:system}.
There are two areas where a model is necessary: determining the extent of the system perturbation, and determining the appropriate response given a particular objective (\eg symmetric flux distribution).
In order to have a fast response (on the order of miliseconds), we must extract specific capabilities of general-purpose system codes to a surrogate model.
Thus, we have adopted current state-of-the-art neural network libraries to build surrogate models.

\subsection{Previous applications of Neural Networks}

Previous work in the applications of Neural Networks (NNs) to Nuclear Engineering problems is summarized.
Several authors have focused on \textit{identification of transients} \cite{Basu1994,Santosh2007,Bartlett2017} such as: LOCA, CR ejection, total loss of off-site power, \etc
Determining the \textit{optimal fuel loading pattern}, with an objective to flattening flux \cite{Sadighi2002} or achieving a particular burnup \cite{Leniau2015}.
A majority of work was focused on providing a \textit{point parameter regression}:
to determine the thermal power \cite{Roh1991};
to predict DNBR using NNs \cite{Kim1997,Lee2003} and hybrid techniques \cite{Zhao2020};
to predict the $k_{eff}$ and maximum power \cite{Mazrou2009}.
Only a single study was found that considered a \textit{multi-dimensional regression} problem.
The work used a NN to predict the transient 3-D power distribution of a theoretical homogeneous cubic reactor \cite{Boroushaki2005}.

\subsection{Objectives}

The literature review indicates that there has been no work in providing a multidimensional regression of a realistic nuclear facility.
Towards achieving our goal of demonstrating autonomous control in the MGEP, we first assess using a neural network surrogate against a well established model.
Thus, this study focused on providing a NN surrogate of the MIT reactor (MITR).
The \pkg{MCNP5} MITR model used in this study has been thoroughly validated \cite{sun2014validation}.
The cross-sectional geometry of the MITR is presented in \cref{fig:mape}.
There are 27 total positions which are filled with fuel elements, aluminum dummies, or experiments.
In this study, a 22-element core is modeled, with other positions occupied by dummies.
The result will be a surrogate model that will accept a control rod (shim blade) position vector, and provide a full-core power distribution.

\section{Surrogate Model}

Our work uses a neural network as a surrogate model. 
Neural networks provide several advantages over traditional machine learning algorithms (SVM, Random Forest, \etc).
Neural networks are under active development and the underlying algorithms are continuously optimized for deployment on various computing architectures.
There are powerful open-source libraries that abstract the development process and allow rapid deployment.
Additionally, the architectures of neural networks can be modified to address varying problems (regression of power distribution, or, inversely regression of control rod position).

\begin{figure}
\includegraphics[width=0.9\linewidth]{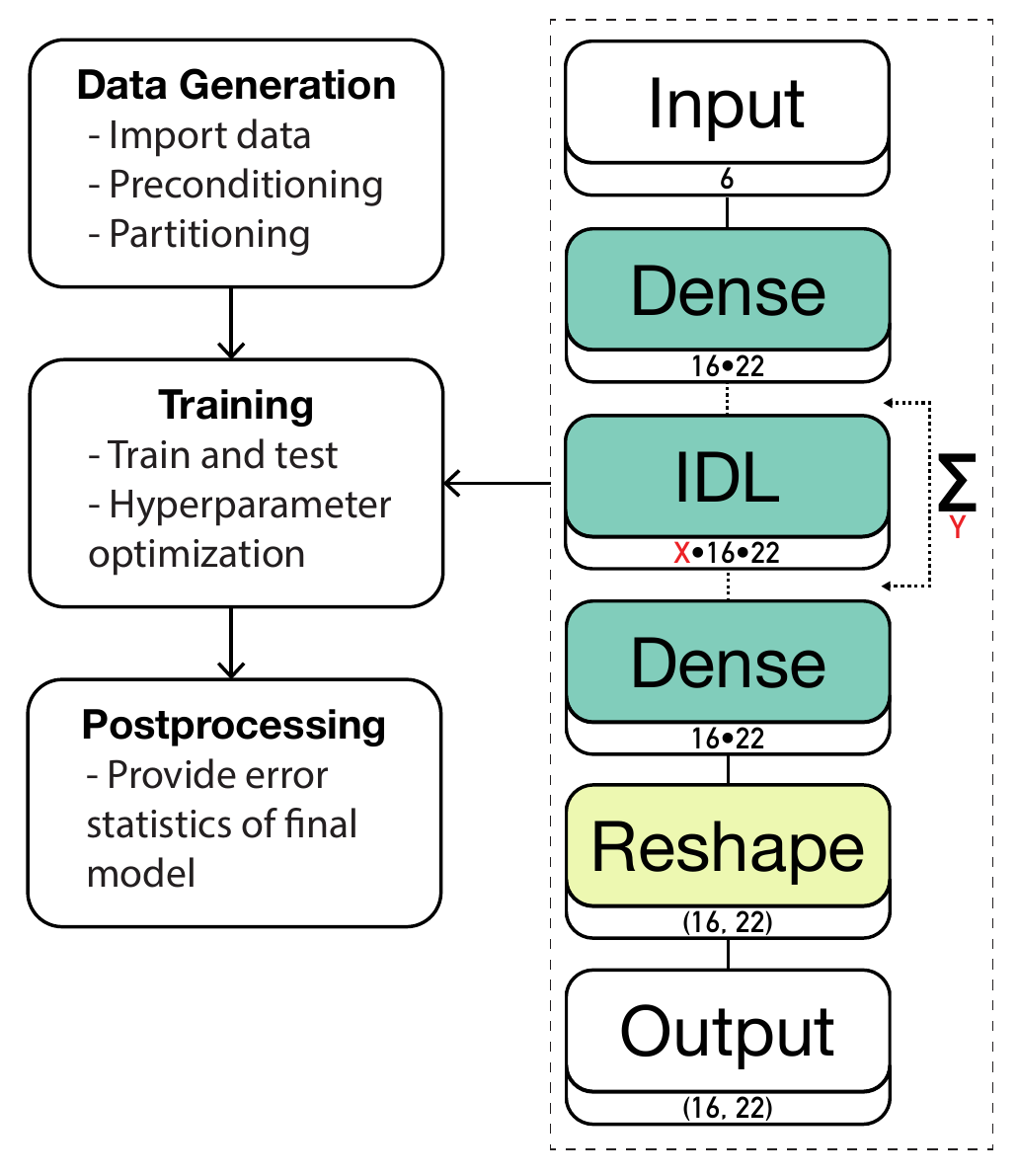}
\caption{\href{https://github.com/a-jd/npsn}{\pkg{NPSN}} package layout. The left-hand graph shows the three major components. The right-hand graph shows the neural network architecture and examples of hyperparameters that are modified during the optimization process (\textcolor{hypred}{\textbf{\htxt{X}}}: IDL layer shape, \textcolor{hypred}{\textbf{\htxt{Y}}}: number of layers)}\label{fig:arch}
\end{figure}

\href{https://github.com/a-jd/npsn}{\pkg{NPSN}} is the package developed to support this work and is available online at \href{https://github.com/a-jd/npsn}{github.com/a-jd/npsn}.
The major components of \href{https://github.com/a-jd/npsn}{\pkg{NPSN}} is summarized in \cref{fig:arch}.
The preparation of datasets involved preparing multiple permutations of the shim blade heights.
Latin hypercube sampling of 6 heights was used to generate 151 permutations.
For each permutation, an input deck for \pkg{MCNP5} v1.60 is generated and executed.
The output from \pkg{MCNP5} is post-processed to generate power distributions for all 22 elements, with each element further discretized into 16 axial nodes.
Therefore, the input of the NN will be a vector of size $(6)$, and the output will be a matrix of size $(16,22)$.

The NN architecture is dependent on the type of problem and data structure -- there is no precise prescription for the structure.
The general structure implemented is presented in \cref{fig:arch}.
The \htxt{Dense}\footnote{Details on layer functions available at \url{https://keras.io} \cite{chollet2015keras}.} layer is used for all intermediate connections between the input and the output. 
There is an Intermediate Dense Layer (\htxt{IDL}), which consists of a variable quantity of \htxt{Dense} layers and a variable shape.
To arrive at an optimal configuration systematically, we have implemented a meta-learning procedure.
The structure and hyperparameters of the neural network are optimized based on a Tree of Parzen Estimator \cite{Bergstra2013}. 
Several hyperparameters such as the batch size, IDL number of layers, IDL shape of layers, IDL activation function, network loss function, \etc, were probed.

Post-processing is an important step in determining the viability of the model.
This is achieved by evaluating the error in providing a regression of ``unseen'' input datasets (known as test data).
If the error of the test datasets matches the error in evaluating training data, the model has good \textit{generalization} and can be expected to perform accurately for novel inputs.
The error is defined as the mean absolute percentage error (MAPE),
\begin{equation}
\varepsilon_{i,j} = \frac{1}{N}\sum_{k=1}^N\left|\frac{\hat{x}_{i,j,k}-x_{i,j,k}}{x_{i,j,k}}\right|~,
\end{equation}
where $\hat{x}_{i,j,k}$ is the predicted power and $x_{i,j,k}$ is the \pkg{MCNP5} power. 
The subscript $i$ represents the element node, $j$ the axial node, $k$ a particular permutation of the shim blade heights, and $K$ is the total number of permutations. 
The total number of summations $N$, depends on the averaging mode.
If we seek element-wise error, core-wise error, $\varepsilon_{i,j}$, $N=K$.
Additional summations can lead to, \eg total error $\varepsilon$, $N=22\cdot 16\cdot K$.
In addition to accuracy, we also test for precision of the model.
The precision is an important consideration as large variances in model outcome could lead to an unstable system.
The precision is quantified by the standard deviation of the error amongst the test datasets,
\begin{equation}
\sigma_{i,j}= \sqrt{\frac{1}{N}\sum_{k=1}^N\left(\varepsilon_{i,j,k}-\varepsilon_{i,j}\right)^2}~,
\end{equation}
where $\varepsilon_{i,j,k}$ is the error before averaging over $k$.

\subsection{Optimization}

\begin{figure}
\includegraphics[width=0.88\linewidth]{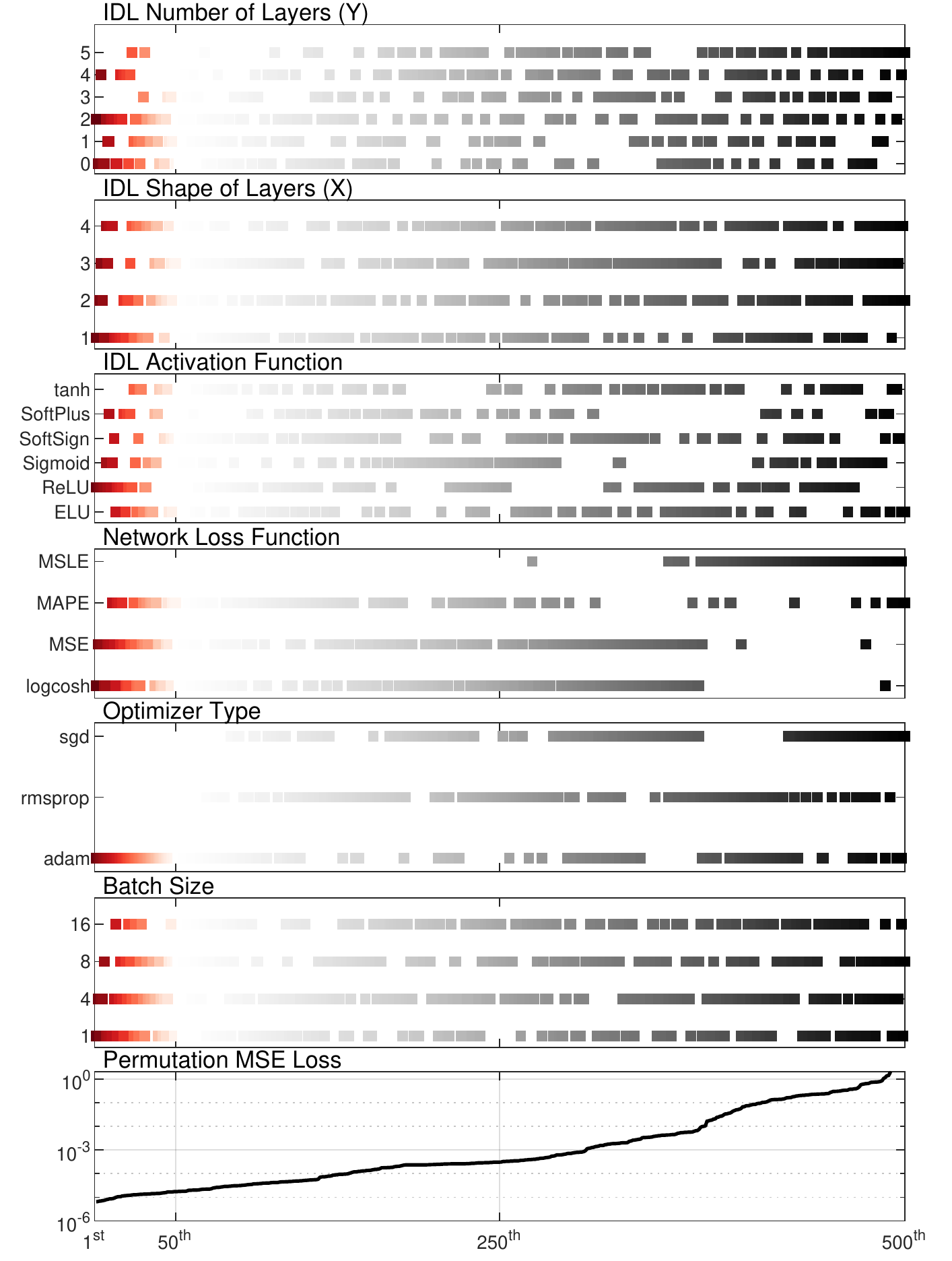}
\caption{Outcome of 500 iterations during optimization. 
	Each plot represents a particular hyperparameter that was modified. 
	The x-axis represents permutations sorted in ascending loss. 
	The top 50 model configurations are highlighted in red.}
	\label{fig:optim_results}
\end{figure}

The optimization process provides a useful guideline for selection of hyperparameters.
In this work, the optimization process involved 500 iterations, presented in \cref{fig:optim_results}.
There are some clear patterns that lead to a more successful model.
A quantity and shape of IDL does not translate to a better model. Thus, an excessively large model is detrimental to performance.
There is a benefit in using the \textit{logcosh} loss function when evaluating the neural network, and using the \textit{ReLU} IDL activation function.
There is a significant benefit in using the \textit{adam} optimizer during training.
On the opposite end of the spectrum, apart from the network loss function, there are no patterns noted that guarantee a poor model.
Therefore, a systematic optimization process is vital in building a neural network.

\section{Results}

\begin{figure}
\includegraphics[width=\linewidth]{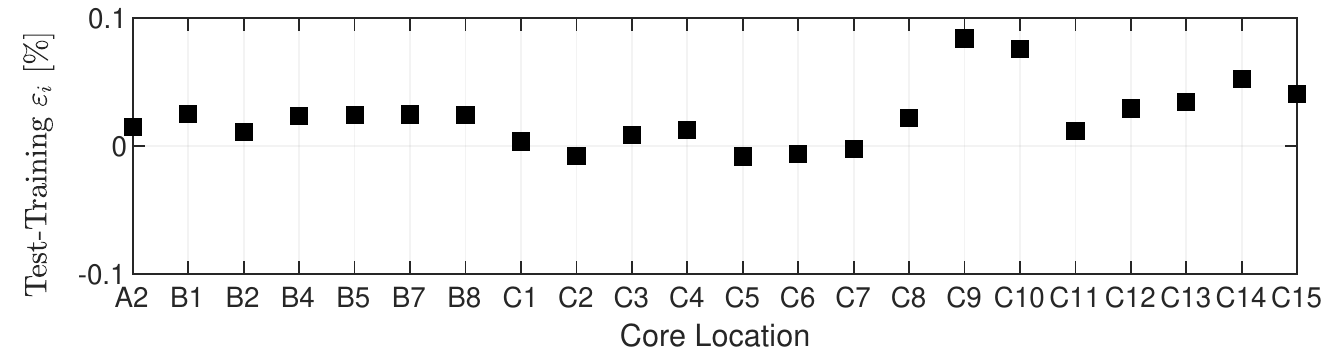}
\caption{Difference between test \& training set element-wise error.
		The training set contains data that the surrogate model has been exposed to and used for tuning neural network parameters.
		The test set contains data that the model has not been exposed to.
		A large discrepancy would indicate that the model is unable to provide a regression for unseen data.}
		\label{fig:generr}
\end{figure}

This section will detail performance metrics of the optimized model.
We are interested in arriving at a model that generalizes well, while minimizing error and noise.
The model had a less than 0.1 \% difference in test \vs training $\varepsilon_{i}$, presented in \cref{fig:generr}.
Therefore, we can expect the model to provide a good regression of unseen input data.
However, this assumption is invalid if the input range exceeds that of the training dataset (\ie if a shim blade is in an anomalous position).

\begin{figure}
\includegraphics[height=3.7cm]{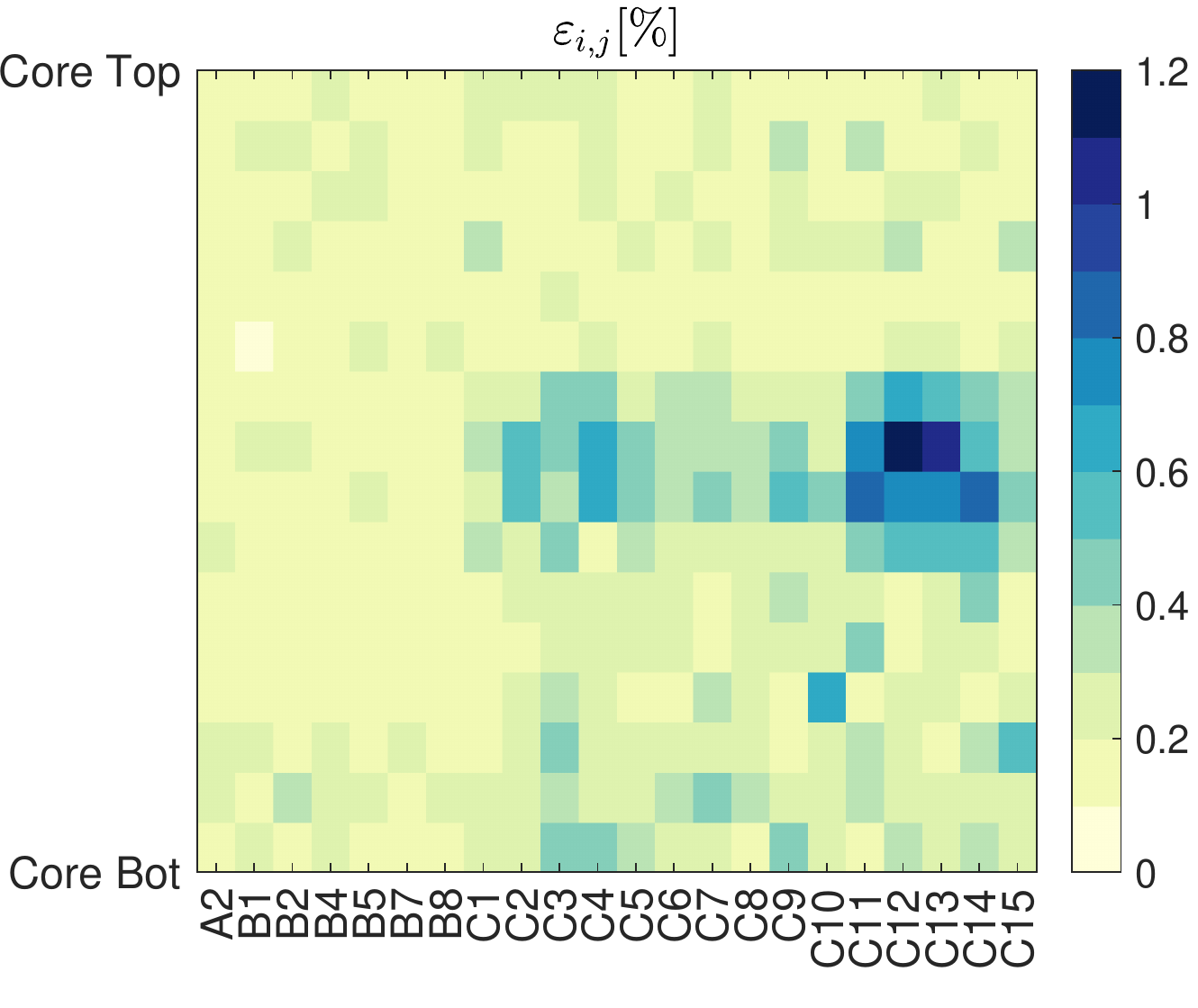} \includegraphics[height=3.7cm,trim=2.1cm 0 0 0,clip]{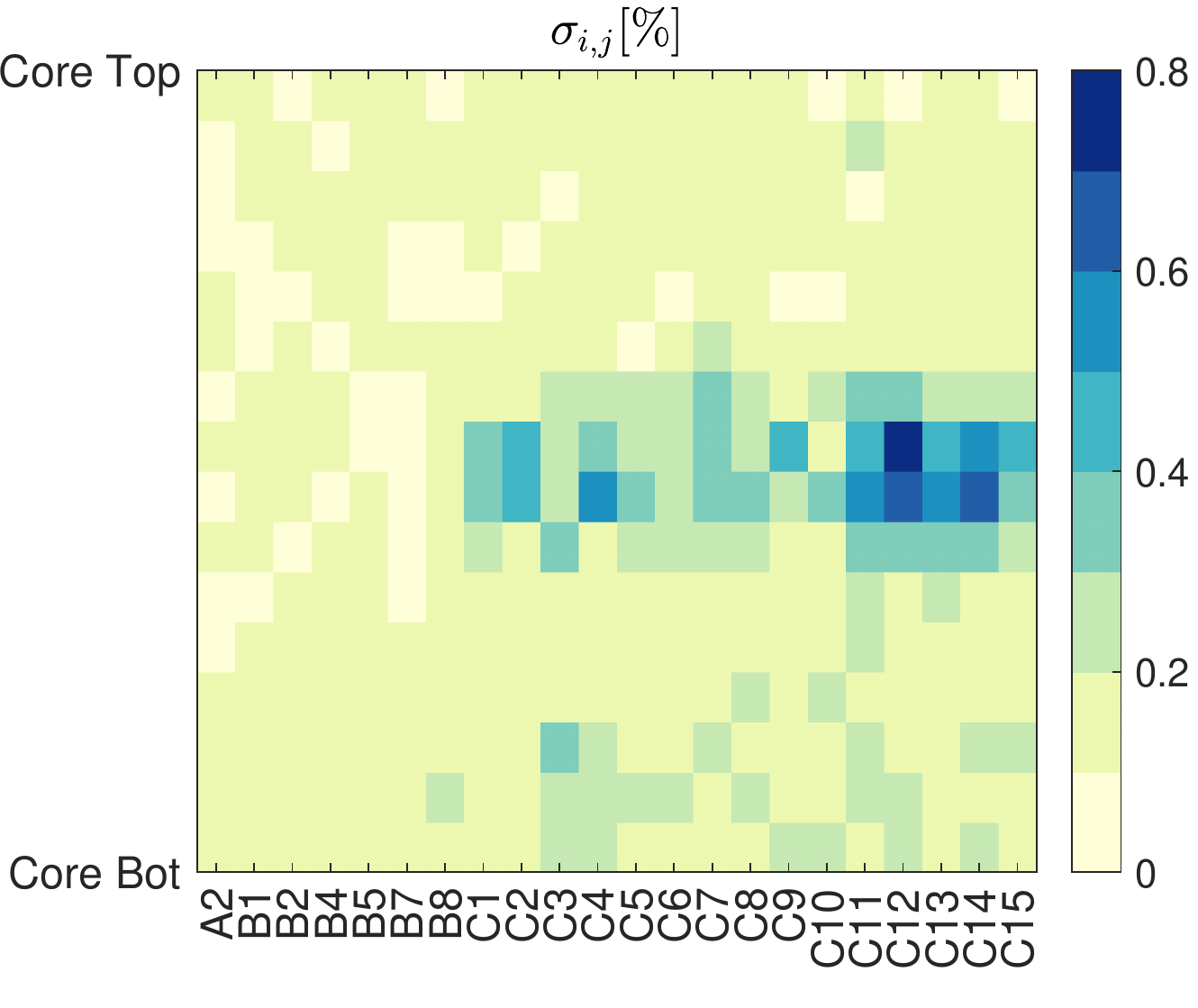}
\caption{Left: Evaluation of the error function across entire test dataset. Right: Evaluation of the error function standard deviation across the entire test dataset.}
\label{fig:errdist}
\end{figure}

\begin{figure}
\includegraphics[width=0.95\linewidth]{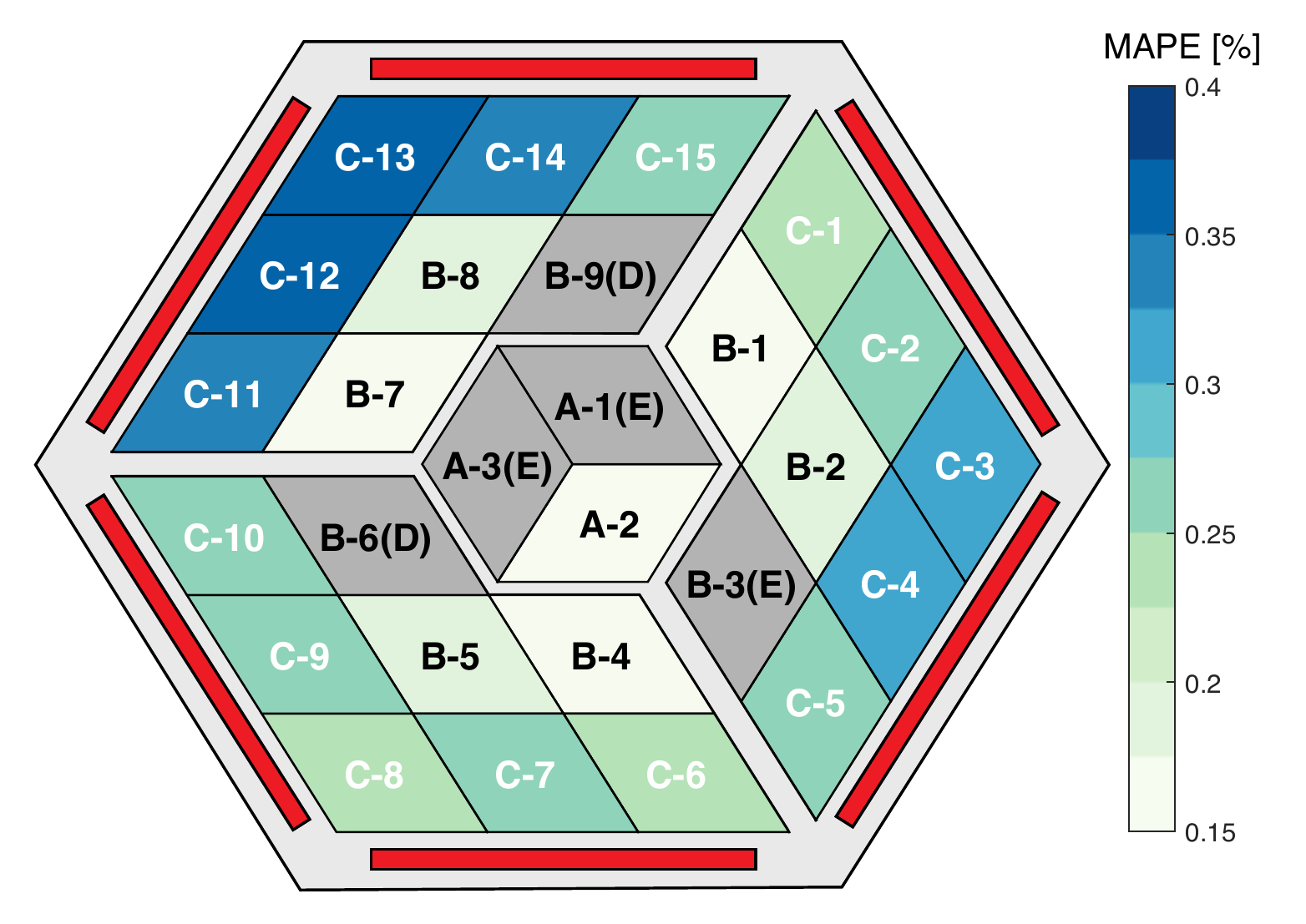}
\caption{Element-wise error ($\varepsilon_{i}$) across all test sets. 
	The red bars indicate the approximate shim blade (control rod) locations in the MITR.
	Gray elements with suffix (D) indicate empty fuel element positions.
	Gray elements with suffix (E) indicate empty in-core experiment positions.
	Empty positions are modeled as dummy aluminum elements.}
	\label{fig:mape}
\end{figure}

The spatial distribution of error and its standard deviation is discussed next.
The values of $\varepsilon_{i,j}$ and $\sigma_{i,j}$ are presented in \cref{fig:errdist}.
There is a coherent feature in the spatial distribution of both parameters: the geometric area towards the center of the core and C-ring elements have larger values of $\varepsilon_{i,j}$ and $\sigma_{i,j}$.
The C-ring is the outermost ring of the MITR core.
In fact, the shim blades are inserted towards the outer region of the core.
Furthermore, the tip of the blades lie towards the center of the core nominally (\ie centroid of sampled heights).
Therefore, the larger values of $\varepsilon_{i,j}$ and $\sigma_{i,j}$ correspond to spatial locations which experience the greatest perturbations from shim blade movement.
This outcome is reasonable as we would expect spatial regions where the power distribution is relatively static, with respect to shim blade movement, to be predicted with far greater accuracy.
The spatial dependence of the error is demonstrated in \cref{fig:mape}.

The magnitude of error and its standard deviation is discussed next.
The error ranges from 0.10-1.16 \%, over 31 \textit{test} datasets.
The corresponding standard deviation ranges from 0.06-0.77 \%. 
Since this is a first-of-its-kind study, there is no available literature to contrast to.
However, the maximum error plus maximum standard deviation (1.93 \%) falls below the experimental uncertainty of the neutron detectors we will use ($\approx2$ \%).
Therefore, our study shows that a NN is a viable surrogate to use in conjunction with experimental data.

\begin{figure}
\includegraphics[width=\linewidth]{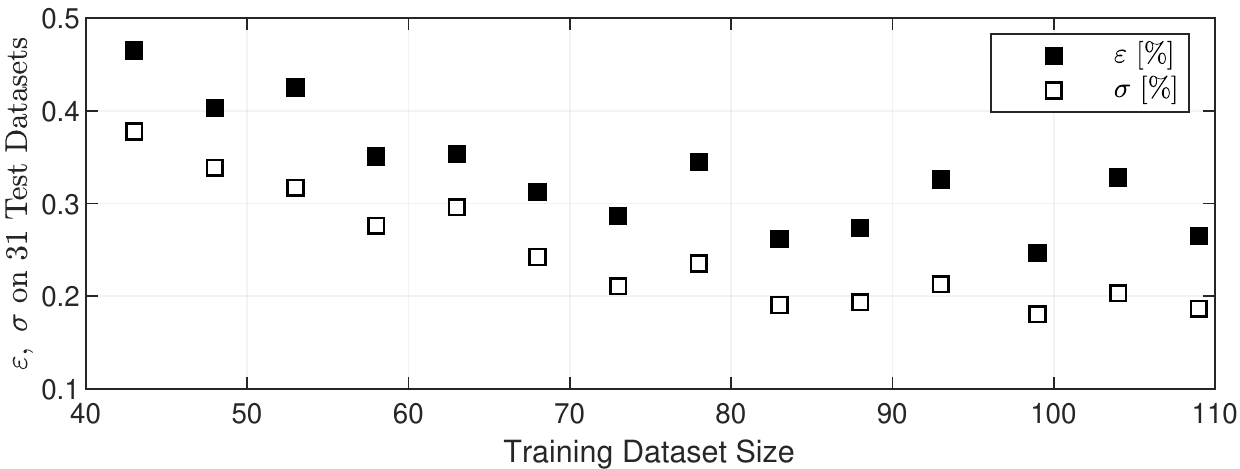}
\caption{Variation in model performance as a function of total dataset size. 
	The composition of test sets is kept constant, while the training sets vary.
	}
	\label{fig:mape_vs_data}
\end{figure}

As the training data (generated by \pkg{MCNP5}) is limited by computational resources, it is interesting to determine if sufficient data has been generated.
The number of training sets \vs $\varepsilon$ and $\sigma$ is presented in \cref{fig:mape_vs_data}.
It is apparent that, initially, the error and standard deviation is decreasing with respect to training dataset size.
After the training dataset size is greater than 80, there improvement saturates.
Thus, our work shows that approximately 100 datasets are necessary to appropriately train and test a surrogate neural network.

Lastly, the computational runtime to provide a regression is highlighted.
During the \pkg{MCNP5} data generation process, 151 datasets were generated.
Each dataset took $\approx\SI{12}{\hour}$ on a 32-core processor to achieve satisfactory statistics.
In contrast, the time required to provide a regression using the NN surrogate is $\approx\SI{5}{\micro\second}$ (using a single NVIDIA TITAN RTX GPU).
The runtime can be reduced further if we optimize the NN compilation using \href{https://github.com/tensorflow/tensorrt}{\pkg{TensorRT}}.
Therefore, the surrogate model will not be the limiting component in the overall system response.

\section{Concluding Remarks}
This work focuses on establishing the capability of neural networks to provide an accurate and precise multi-dimensional regression of a nuclear reactor's power distribution.
The results indicate that neural networks are an appropriate choice for surrogate models to implement in an autonomous reactor control framework.
The MAPE across all test datasets was $<$ 1.16 \% with a corresponding standard deviation of $<$ 0.77 \%.
The error is low, considering that the node-wise fission power can vary from \SI{7}{\kilo\watt} to \SI{30}{\kilo\watt} across the core.
This work also provides guidance for best practices in network architecture, hyperparameter selection and dataset size.

The code used in this work is available online as an open-source python package, \href{https://github.com/a-jd/npsn}{\pkg{NPSN}}.
The package is written to abstract the process of importing and pre-conditioning data, optimizing the neural network architecture, and post-processing.
An example of the code syntax for the end user:

\begin{lstlisting}
import npsn

# Define dataset directory
data_dir = '~/some/data_location'
# Define model name (for output file label)
proj_nm = 'npsn_surrogate'

# Define number of control blades
n_x = 6
# Define nodalization of power distribution
n_y = (16, 22)  #(axial_nodes, fuel_locations)

# Train neural network without optimization
npsn.train(proj_nm, data_dir, n_x, n_y)
# Or with optimization
npsn.train(proj_nm, data_dir, n_x, n_y, max_evals=100)
# Post-process to quantify error
npsn.post(proj_nm)
\end{lstlisting}

\section{Acknowledgments}
This work is supported by DOE NEUP Award Number: {DE-NE0008872}.
\bibliographystyle{ans}
\bibliography{bibliography}
\end{document}